\documentclass[12pt]{article}

\newcommand\eq[1] {(\ref{#1})}

\newcommand{\bfm}[1]{\mbox{\boldmath ${#1}$}}
\newcommand{\nonum}{\nonumber \\}
\newcommand{\beqa}{\begin{eqnarray}}
\newcommand{\eeqa}[1]{\label{#1}\end{eqnarray}}
\newcommand{\beq}{\begin{equation}}
\newcommand{\eeq}[1]{\label{#1}\end{equation}}

\newcommand{\lang}{\langle}
\newcommand{\rang}{\rangle}

\newcommand{\Ga}{\alpha}
\newcommand{\Gb}{\beta}
\newcommand{\Gd}{\delta}

\newcommand{\Gve}{\varepsilon}

\newcommand{\Gl}{\lambda}
\newcommand{\Gn}{\eta}
\newcommand{\Gm}{\mu}
\newcommand{\Gv}{\nu}

\newcommand{\BGe}{\bfm\epsilon}
\newcommand{\BGve}{\bfm\varepsilon}

\newcommand{\BGm}{\bfm\mu}

\def\Bx{{\bf x}}

\def\BC{{\bf C}}

\def\BI{{\bf I}}

\def\BP{{\bf P}}
\def\BQ{{\bf Q}}

\def \ba {\begin{array}}
\def \ea {\end{array}}
\newtheorem {Thm} {Theorem} [section]

\newtheorem {Adef} [Thm] {Definition}

\newtheorem {Arem} [Thm] {Remark}

\newtheorem {Aexa} [Thm] {Example}

\newtheorem {Anot} [Thm] {Notation}

\def \refe #1.{(\ref{#1})}
\def \reff #1.{figure~\ref{#1}}
\def \refs #1.{section~\ref{#1}}
\def \refss #1.{subsection~\ref{#1}}
\def \refD #1.{Definition~\ref{#1}}
\def \refT #1.{Theorem~\ref{#1}}
\def \refL #1.{Lemma~\ref{#1}}
\def \refC #1.{Corollary~\ref{#1}}
\def \refP #1.{Proposition~\ref{#1}}
\def \refR #1.{Remark~\ref{#1}}
\def \refE #1.{Example~\ref{#1}}
\def \refN #1.{Notation~\ref{#1}}

\usepackage{graphics}
\usepackage{amssymb}
\begin{document}
\vspace{-1in}
\title{Realizability of metamaterials with prescribed electric permittivity and magnetic permeability tensors}
\author{Graeme W. Milton \\
\small{Department of Mathematics, University of Utah, Salt Lake City UT 84112, USA}}
\date{}
\maketitle
\begin{abstract}
We show that any pair of real symmetric tensors $\BGve$ and $\BGm$ can be realized
as the effective electric permittivity and effective magnetic permeability of
a metamaterial at a given fixed frequency.
The construction starts with two extremely low loss metamaterials, 
with arbitrarily small microstructure, whose existence 
is ensured by the work of Bouchitt{\'e} and Bourel and Bouchitt\'e and Schweizer,
one having at the given frequency a 
permittivity tensor with exactly one negative eigenvalue, and a positive permeability
tensor, and the other  having a positive permittivity tensor, and a permeability
tensor having exactly one negative eigenvalue.
To achieve the desired effective properties these materials are laminated together in 
a hierarchical multiple rank laminate structure, with widely separated length scales,
and varying directions of lamination, but with the largest length scale still much shorter
than the wavelengths and attenuation lengths in the macroscopic effective medium.
\end{abstract}
\vskip2mm


\vskip2mm
\section*{Introduction}
Isotropic materials with negative values of permittivity $\Gve$ and/or permeability $\Gm$ (with low loss) 
have some fascinating properties. The colors of some stained glass windows are
due to a resonance effect caused by metal spheres, which have a
negative permittivity, embedded in the glass \cite{MaxwellGarnett:1904:CMG}. 
Veselago \cite{Veselago:1967:ESS} found that a slab of material with $\Gve=\Gm=-1$
at a given frequency would have a negative refractive index and act as a lens. 
Nicorovici, McPhedran and Milton \cite{Nicorovici:1994:ODP} found that a cylindrical shell 
with permeability $\Gve=-1$ would, in the quasistatic limit and for TM waves, magnify the core
material and produce a perfect image of a line dipole in the limit as the loss 
went to zero. Pendry \cite{Pendry:2000:NRM} made the remarkable assertion
that the Veselago lens would behave as a superlens, not limited by diffraction, providing
perfect images of point sources, not just in the quasistatic limit,
and now the validity of this assertion is beyond doubt provided one adds an
infinitesimal loss to the lens: see, for example, 
\cite{Cummer:2003:SCS,Shvets:2003:PAM,Merlin:2004:ASA,Podolskiy:2005:NSS,Milton:2005:PSQ}.
Polarizable dipoles can
be cloaked by such lenses \cite{Milton:2006:CEA,Nicorovici:2007:OCT,Milton:2006:OPL,Bruno:2007:SCS,Bouchitte:2010:CSO}, and 
larger objects can be cloaked by embedding the matching ``antiobject'' in the lens \cite{Lai:2009:CMI}.
However any reasonable losses in the material can almost destroy these effects.
From a practical viewpoint losses are always present and even extremely tiny losses can prevent the functioning
of the Pendry superlens if the wavelength is short compared with the lens 
thickness \cite{Podolskiy:2005:NSS,Kuhta:2007:FFI}. However the negative refractive index property
is robust: materials with a negative refractive index were manufactured by Shelby, Smith and Schultz \cite{Shelby:2001:EVN}.

Other remarkable applications are associated with anisotropic materials having permittivity
tensors $\BGve$ and permeability tensors $\BGm$ with $\BGve=\BGm$. It has long been known
that the time harmonic Maxwell's equations are invariant under curvilinear coordinate transformations \cite{Post:1962:FSE}
and Dolin \cite{Dolin:1961:PCT} realized 
that the transformed equations can be viewed as a solution of Maxwell's equations in a new
material. Using such transformations empty space with $\Gve=\Gm=1$ is transformed in a 
material with $\BGve(\Bx)=\BGm(\Bx)$: thus Dolin
recognized that certain inhomogeneous inclusions with
$\BGve(\Bx)=\BGm(\Bx)$ are equivalent to empty space and are thus invisible. 
Pendry, Schurig and Smith \cite{Pendry:2006:CEM}
went one step further and, similar to what Greenleaf, Lassas and Uhlmann \cite{Greenleaf:2003:ACC} had done 
in the context of the conductivity equations, found that by using a transformation
which maps a point to a sphere one could create a cloak having $\BGve(\Bx)=\BGm(\Bx)$
which completely shields an object, making it invisible to a fixed frequency of incident radiation.
Rigorous mathematical justifications of this cloaking, and its generalizations, have been given 
\cite{Greenleaf:2007:FWI,Weder:2008:RAH,Kohn:2009:CCV}, and various approximate cloaks have been 
proposed and some have been physically
constructed \cite{Schurig:2006:MEC,Cai:2007:OCM,Cai:2007:NMC,Li:2008:HUC,Liu:2009:BGP,Valentine:2009:OCD}.
This and other types of cloaking are surveyed in reviews \cite{Alu:2008:PMC,Greenleaf:2009:CDE}.
The growing field of transformation optics uses coordinate transformations
to achieve unusual effects, such as rotators \cite{Chen:2007:TMR},
concentrators \cite{Rahm:2008:DEC}, wormholes \cite{Greenleaf:2008:EWV},
novel superlenses \cite{Kildishev:2007:IMH, Milton:2008:SFG,Yan:2008:CSC,Tsang:2008:MPL}
but requires one to tailor materials with prescribed tensors $\BGve$ and $\BGm$.
In the geometric optics limit one can use transformations of the refractive index to achieve
cloaking \cite{Leonhardt:2006:OCM,Leonhardt:2009:BIN} and novel Eaton lenses \cite{Tyc:2008:TSO}.

Extreme values of $\BGve$ and $\BGm$, near zero or infinity, also have striking applications, leading to new types of 
circuits \cite{Engheta:2005:CEO,Milton:2009:EC,Milton:2010:HEC} and tunneling of energy through 
narrow channels \cite{Silveirinha:2006:TEE}.

Despite these tantalizing results, many of which are summarized in the book of Cai and Shalaev \cite{Cai:2010:OM},  
to my knowledge no one has either experimentally, numerically or theoretically
shown that any given pair of prescribed symmetric real tensors $\BGve$ and $\BGm$,
and in particular given tensor pairs with  $\BGve=\BGm$, including isotropic materials with $\Gve=\Gm=-1$,
can be almost realized at a fixed frequency. (Isotropic materials with negative real permittivity and 
permeability have been realized \cite{Efros:2004:DPC}, but it is unclear if they can be realized with
$\Gve=\Gm=-1$.) This paper accomplishes that goal from a theoretical
perspective. It shows there are no hidden constraints restricting the possible real pairs $(\BGve,\BGm)$ that can
exist at one given frequency.
The construction uses hierarchical microstructures with structure on many widely separated length scales,
and component metamaterials with arbitrarily low loss. It seems likely that the same effective properties could be achieved
with more realistic designs, but it is doubtful that their properties could be calculated analytically.

\section*{The starting materials and Tartar's formulae for the effective tensors of laminates}

In this paper we assume an idealized world where the continuum, and Maxwell's equations,
extend down to arbitrarily small length scales and we assume the existence of two metamaterials. Material $A$ has 
arbitrarily small microstructure and an effective permittivity tensor 
with a bounded real part $(\BGve^A)'$ which is diagonal, 
with exactly one negative and 
two positive diagonal elements and an effective permeability tensor 
with a real part $(\BGm^A)'$ which is not necessarily diagonal but bounded and strictly positive definite, 
i.e. there exist constants $\Gb_A>\Ga_A>0$ such that
\beq \Gb_A\BI>(\BGm^A)'>\Ga_A\BI.
\eeq{1.2b}
The imaginary parts $(\BGve^A)''$ and $(\BGm^A)''$ of these two tensors are assumed to be vanishing small,
but non-zero, i.e. there exist constants $\Gn>\Gv>0$ such that
\beq \Gn\BI>(\BGve^A)''>\Gv\BI, \quad \Gn\BI>(\BGm^A)''>\Gv\BI, \eeq{1.2ba}
where the loss parameters $\Gn$ and $\Gv$ are vanishingly small. [Strictly speaking we should
talk about a sequence of materials $A$, parameterized by $\Gn$ and consider what happens
in the limit as $\Gn\to 0$, when $(\BGve^A)'$ and 
$(\BGm^A)'$ do not depend on $\Gn$, while $\Gv$ and the scale and material constants 
of the microstructure within the metamaterial, do depend on $\Gn$.] Material $B$ has 
arbitrarily small microstructure and an 
an effective permittivity tensor with a real part $(\BGve^B)'$ which is strictly positive definite, i.e. satisfying the bounds
\beq \Gb_B\BI>(\BGve^{B})'>\Ga_B\BI, \eeq{1.9}
for some choice of $\Gb_{B}>\Ga_{B}>0$, and with an effective permeability with a bounded real part $(\BGm^B)'$ which is diagonal, with exactly one negative and two positive diagonal elements. The imaginary parts  
$(\BGve^B)''$ and $(\BGm^B)''$ of these two tensors are assumed to be vanishing small but non-zero, 
i.e. to satisfy the bounds
\beq \Gn\BI>(\BGve^B)''>\Gv\BI, \quad \Gn\BI>(\BGm^B)''>\Gv\BI, \eeq{1.9a}
where, without loss of generality $\Gn$ and $\Gv$ are the same parameters as appear
in \eq{1.2ba}. [Again strictly speaking we should consider a sequence of materials $B$, 
parameterized by $\Gn$ and consider what happens in the limit as $\Gn\to 0$.]

Material $A$ could be a metamaterial comprised of a cubic lattice of 
well separated cubes, where each cube has a microstructure of highly   
conducting rods aligned in the $x_1$ direction. A rigorous
mathematical proof that such a material has the desired effective properties,
with a more precise description of the needed microgeometry, has been given by Bouchitt{\'e} and Bourel
\cite{Bouchitte:2010:HFM}.
Based on the results of Pendry et.al.
\cite{Pendry:1996:ELF} and Bouchitt{\'e} and Felbacq \cite{Bouchitte:1997:HSP}
one might think that an periodic array of highly
conducting thin rods aligned in the $x_1$ direction would suffice for material $A$. However, Bouchitt{\'e} and
Felbacq \cite{Bouchitte:2006:HWP} show that in a finite sample of such a material the associated effective equations 
can have a non-local behavior.

Material $B$ could be a metamaterial comprised of a periodic lattice of highly conducting
split ring resonators with axes aligned in the $x_1$ direction with one rod per unit cell. 
The split rings behave like polarizable magnetic dipoles and if one is just above resonance
these can have a negative permeability in the $x_1$ direction. This means of creating
artificial magnetism was discussed by Schelkunoff and Friis
\cite{Schelkunoff:1952:ATP}, and although their formula showed a negative
permeability in the $x_1$ direction, they did not draw attention to this fact. The microstructure
was rediscovered by Pendry et.al. \cite{Pendry:1999:MCE}, who realized the significance of the negative permeability. A rigorous
mathematical proof that such a material has the desired effective properties,
has been given by Bouchitt\'e and Schweizer \cite{Bouchitte:2010:HME}, following earlier work of 
\cite{Zhikov:2000:EAT,Zhikov:2004:GSE,Bouchitte:2004:HNR,Felbacq:2005:TMM,Kohn:2008:MHM} 
(see also the introduction of \cite{Smyshlyaev:2009:PLE}).
Bouchitt\'e and Schweizer also prove that one can obtain isotropic effective magnetic permeability tensors
with a negative permeability.

We explore what effective permittivity and permeability tensors can be generated
from composites of these two metamaterials $A$ and $B$ and all rotations of them
within classical homogenization in the 
limit as $\Gn$ tends to zero. The first step will be to recover the result
of Bouchitt{\'e} and Bourel \cite{Bouchitte:2010:HFM} 
that a material with any real symmetric permittivity tensor $\BGve_*$ can be approximately achieved.
Then the same argument applies to show that a material with any real symmetric permeability 
tensor $\BGm_*$ can be approximately achieved. Finally by combining these two results we show that any pair
of real symmetric tensors $(\BGve_*,\BGm_*)$ can be approximately achieved.

The composites we consider are multiple rank 
laminates. This class of composite was first introduced by Maxwell \cite{Maxwell:1954:TEM} and has been
extensively studied in the homogenization literature: see, for example, \cite{Milton:2002:TOC} and references
therein. They are obtained by laminating 
the starting (rank 0) materials 
together on an extremely small length scale to obtain rank 1 laminates
and then laminating these rank 1 laminates with other rank 1 or rank 0 laminates 
on a much larger, but still extremely 
small length scale, using a possibly different direction of lamination, 
to obtain rank 2 laminates and so forth.  The advantage of considering this class of
composites is that their effective tensors can be explicitly calculated, in the
limit in which the ratio of successive length scales approaches infinity. The 
bounds \eq{1.2ba} and \eq{1.9a} ensure that one can use reiterated homogenization
theory: to calculate the effective tensor in the limit of widely separated scales
of a multiple rank laminate which is a simple laminate of two (possibly multiple) rank laminates $C$ and $D$
one can replace $C$ and $D$ by homogeneous materials with material tensors the same
as the effective tensors of $C$ and $D$.  

The use of classical homogenization, except inside the metamaterials $A$ and $B$, 
should be valid provided, when we replace materials $A$ and $B$ by homogeneous 
materials with material tensors the same as the effective tensors of $A$ and $B$,
the wavelengths and attenuation lengths within each substructure should be much
larger than the microstructure at that level. For any fixed $\Gn>0$ 
this should be ensured by \eq{1.2ba} and \eq{1.9a},
in the limit in which the overall microstructure tends to zero, and the ratio 
between scales tends to infinity.

If a set of materials are laminated in direction $x_1$
then the formulae obtained by Tartar \cite{Tartar:1979:ECH} 
(using an idea that goes back to Backus \cite{Backus:1962:LWE})
for the effective permittivity and permeability are
\beq \widetilde{\BGve}^*=\lang\widetilde{\BGve}\rangle,\quad
\widetilde{\BGm}^*=\langle\widetilde{\BGm}\rang,
\eeq{1.1}
where $\BGve(x_1)$ and $\BGm(x_1)$ are the local permittivity and permeability
tensors (that could themselves be effective tensors) which only depend on $x_1$ and
the angular brackets denote volume averages and
for any symmetric matrix $\BC$ with elements $c_{ij}$ the matrix $\widetilde{\BC}$ 
is symmetric and has elements
\beq \widetilde{c}_{11}=-1/c_{11},\quad \widetilde{c}_{1k}=c_{1k}/c_{11},\quad
\widetilde{c}_{k\ell}=c_{k\ell}-c_{k1}c_{1\ell}/c_{11},
\eeq{1.2}
for all $k\ne 1$, $\ell\ne 1$. Conversely if a symmetric matrix $\widetilde{\BC}$
is given, then $\BC$ is symmetric with elements
\beq c_{11}=-1/\widetilde{c}_{11},\quad c_{1k}=-\widetilde{c}_{1k}/\widetilde{c}_{11}, \quad
c_{k\ell}=\widetilde{c}_{k\ell}-\widetilde{c}_{k1}\widetilde{c}_{1\ell}/\widetilde{c}_{11},
\eeq{1.2aa}
for all $k\ne 1$, $\ell\ne 1$. 

In the case where $\BGve$ is diagonal,
\eq{1.1} and \eq{1.2} reduce to the familiar harmonic and arithmetic averages
\beq 1/\Gve_{11}^*=\lang 1/\Gve_{11}\rang, \quad \Gve_{kk}^*=\lang\Gve_{kk}\rang\quad k\ne 1,
\eeq{1.2a}
with similar formulae applying when $\BGm$ is diagonal.

In the limit as the imaginary part of $\BGve$ approaches zero then \eq{1.1} implies that
the imaginary part of $\widetilde{\BGve}^*$ will also tend to zero unless 
one is at resonance where the real part of $\lang 1/\Gve_{11}\rangle$
or the real part of $\lang 1/\Gm_{11}\rangle$ vanishes. We will always avoid this
happening. Then the limiting tensors $\BGve^*$ and $\BGm^*$ can be obtained  
by replacing in \eq{1.1} $\BGve$ and $\BGm$ with their real parts, i.e. the vanishingly
small imaginary parts do not effect the effective tensors except at resonance.
From now on we will drop the primes and treat $\BGve$ and $\BGm$ as real, while
remembering that they do have vanishingly small imaginary parts to ensure that we
can apply the rules of classical reiterated homogenization.


\section*{Realizing a material with a desired tensor $\BGve^*$ using lamination}

Now let us explore what effective permittivity tensors $\BGve^*$ can be obtained by hierarchically 
laminating material $A$ with itself (and its rotations by $90^\circ$ about the axes)
in directions parallel to the axes, calculating the effective
tensors at each stage using the harmonic and arithmetic averages \eq{1.2a}. Let us assume the axes have been chosen 
so $\BGve^A$=Diag$[-a,b,c]$, with $a$, $b$ and $c$ all positive. The following remark is helpful:
\vskip2mm
{\bf Remark 1} If $\BGve$=Diag$[-a,b,c]$, with $a$ and $b$ both positive is realizable then so is
$\BGve^*$=Diag$[-a/\Gd,b\Gd,c]$ for any finite $\Gd\ne 0$.
\vskip2mm
To prove this remark, let us laminate Diag$[-a,b,c]$ with the rotated material Diag$[b,-a,c]$ in direction
$x_1$ in volume fractions $f_1$ and $f_2=1-f_1$. The resulting material has effective tensor 
$\BGve^*$=Diag$[-a/\Gd,b\Gd,c]$ with $\Gd=f_1-f_2(a/b)$ taking all values inside the interval between $1$ and  
$-a/b$ as $f_1$ ranges between $0$ and $1$, excepting $\Gd=0$ where one of the eigenvalues of $\BGve^*$
becomes infinite and one is at resonance.
Alternatively, let us laminate these two materials in 
direction $x_2$. The resulting material then has effective tensor 
$\BGve^*$=Diag$[-a/\Gd,b\Gd,c]$ with $\Gd=[f_1-f_2(b/a)]^{-1}$ taking all values outside the interval between $1$ and  
$-a/b$ as $f_1$ ranges between $0$ and $1$. Using one of the two constructions any finite value of $\Gd\ne 0$ is possible.
\vskip2mm
Starting from $\BGve^A$=Diag$[-a,b,c]$, with $a$, $b$ and $c$ all positive and applying Remark 1, we obtain a 
material with effective tensor Diag$[a/\Gd_0,-b\Gd_0,c]$, with $\Gd_0>0$. Applying remark 1 again, we obtain 
a material with effective tensor Diag$[a/\Gd_0,-b\Gd_0/\Gd_1,c\Gd_1]$, and by rotation (and replacing $\Gd_1$ with $1/\Gd_2$)
a material with effective tensor Diag$[a/\Gd_0,c/\Gd_2,-b\Gd_2\Gd_0]$. Laminating these last two materials together in
equal proportions in direction $x_1$ gives a material with effective tensor Diag$[a/\Gd_0,e,g]$, where
\beq e=(c/\Gd_2-b\Gd_0/\Gd_1)/2,\quad g=(c\Gd_1-b\Gd_2\Gd_0)/2.
\eeq{1.4}
Given prescribed values of $e>0$ and $g\ne 0$ we may choose non-zero
\beq \Gd_1=\frac{g\pm\sqrt{g^2-(\Gd_0 bcg/e)}}{c},\quad \Gd_2=\frac{c\Gd_1-2g}{b\Gd_0},
\eeq{1.5}
so that \eq{1.4} is satisfied provided we choose $\Gd_0$ with
\beq 0<\Gd_0<|ge|/(bc), \eeq{1.6}
to ensure that the roots of \eq{1.5} are real. Since \eq{1.6} remains valid if we change the
sign of $e$ we can also realize a material with effective tensor  Diag$[a/\Gd_0,-e,g]$, and 
hence by remark 1, a material with effective tensor  Diag$[-a/\Gd_0,e,g]$. Laminating this material
in direction $x_2$ with Diag$[a/\Gd_0,e,g]$ we can obtain a material with a prescribed 
effective tensor Diag$[h,e,g]$ with $e>0$ provided $\Gd_0$ is chosen sufficiently small to satisfy
\eq{1.6} and the constraint that $\Gd_0<a|h|$ which ensures $h$ lies between 
$a/\Gd_0$ and $-a/\Gd_0$. So any diagonal tensor with non-zero diagonal elements, at least one
of which is positive, is realizable. 

The only case left to treat is when $h$, $e$ and $g$ are
all negative. Let $t$ be bigger that both $-e$ and $-g$. Then
the tensors Diag$[h,e+t,g-t]$ and Diag$[h,e-t,g+t]$ are both realizable
and by laminating them together in direction $x_1$ we see that the tensor
Diag$[h,e,g]$ is also realizable as an effective permittivity tensor for any 
non-zero finite choice of  $h$, $e$ and $g$.

\section*{Realizing any desired pair of tensors $(\BGve^*,\BGm^*)$}

By rotating the material just obtained we see that any symmetric matrix $\BGe^{A*}$, except possibly those which
are singular, is realizable as an effective permittivity tensor. The associated effective permeability tensor $\BGm^{A*}$ 
(in the limit the loss goes to zero) must satisfy the classical homogenization Weiner bounds \cite{Wiener:1912:TMF}
\beq \lang \BGm\rang \geq \BGm^{A*} \geq \lang \BGm^{-1}\rang^{-1},
\eeq{1.7} 
where $\BGm(\Bx)$ is the local permeability tensor, which is locally a rotation of $\BGm^A$.
Since $\BGm^A$ satisfies \eq{1.2b} we conclude that $\BGm^{A*}$ also satisfies the
inequality
\beq \Gb_A\BI>\BGm^{A*}>\Ga_A\BI.
\eeq{1.8}
Let us call $U$ the family of materials thus obtained.

Similarly (because we can interchange the roles of $\BGve$ and $\BGm$) using material $B$, and its
rotations, any symmetric matrix $\BGm^{B*}$, except possibly those which
are singular, is realizable as an effective permeability tensor, and the associated effective permittivity
$\BGve^{B*}$ satisfies the bounds
\beq  \Gb_B\BI>\BGve^{B*}>\Ga_B\BI. \eeq{1.10}
We let $V$ denote this family of materials. 

Now we are free to take any set of materials in the set $W=U\cup V$ and laminate them together in direction $x_1$ to form
a larger set of materials $LW$ containing $W$. By an abuse of notation we let $W$ also denote the set of tensor 
pairs $(\widetilde{\BGve}^*,\widetilde{\BGm}^*)$ deriving from effective tensors $({\BGve}^*,{\BGm}^*)$ of materials
in $W$, and we let $LW$ also denote  the set of tensor 
pairs $(\widetilde{\BGve}^*,\widetilde{\BGm}^*)$ deriving from effective tensors $({\BGve}^*,{\BGm}^*)$ of materials
in $LW$. Now \eq{1.1} implies the tensor pairs in $LW$ correspond to all convex combinations of the tensor pairs in $W$,
except for possibly those tensor pairs for which either $\widetilde{\Gve}^*_{11}=0$ or $\widetilde{\Gm}^*_{11}=0$ where one is at resonance. So aside from a set of measure zero the
set of tensor pairs $LW$ is a convex set whose convex hull is characterized by the Legendre transform:
\beq f(\BP,\BQ)=\inf_{(\widetilde{\BGve}^*,\widetilde{\BGm}^*)\in W}\sum_{i=1}^3\sum_{j=1}^3
p_{ij}\widetilde{\Gve}^*_{ij}+q_{ij}\widetilde{\Gm}^*_{ij},
\eeq{1.11}
as a function of the symmetric matrices $\BP$ and $\BQ$. If we can show that $f(\BP,\BQ)$ is minus infinity
for every choice of $\BP$ and $\BQ$, not both zero, then we can conclude that pair of real symmetric 
matrices $(\widetilde{\BGve}^*,\widetilde{\BGm}^*)$ is approximately realizable.

First observe that
\beq f(\BP,\BQ)=\min\{f_A(\BP,\BQ),f_B(\BP,\BQ)\},
\eeq{1.12}
where
\beqa 
f_A(\BP,\BQ)=\inf_{(\widetilde{\BGve}^{A*},\widetilde{\BGm}^{A*})\in U}\sum_{i=1}^3\sum_{j=1}^3
p_{ij}\widetilde{\Gve}^{A*}_{ij}+q_{ij}\widetilde{\Gm}^{A*}_{ij}, \nonum
f_B(\BP,\BQ)=\inf_{(\widetilde{\BGve}^{B*},\widetilde{\BGm}^{B*})\in V}\sum_{i=1}^3\sum_{j=1}^3
p_{ij}\widetilde{\Gve}^{B*}_{ij}+q_{ij}\widetilde{\Gm}^{B*}_{ij}.
\eeqa{1.13}
Since almost every matrix $\BGve^{A*}$ is realizable as a permittivity tensor, it follows that almost every matrix
$\widetilde{\BGve}^{A*}$ is realizable: given $\widetilde{\BGve}^{A*}$ we realize, or almost realize, the 
tensor $\BGve^{A*}$ whose elements are given according to \eq{1.2aa}. In particular we can realize a tensor
$\widetilde{\BGve}^{A*}\approx -\Gl\BP$, for any value of $\Gl$ no matter how large. The associated
permeability tensor $\BGm^{A*}$ satisfies \eq{1.8}, which implies
\beq \Gb_A>\Gm^{A*}_{ii}>\Ga_A, \quad \Gb_A>|\Gm^{A*}_{ij}|~~~{\rm for~all~}i\ne j, \eeq{1.14}
where the latter condition follows from the positivity of the determinant of the
associated $2\times 2$ submatrix of $\BGm^{A*}$. It follows that the elements of $\widetilde{\BGm}^{A*}$, given
according to \eq{1.2}, satisfy the bounds
\beq |\widetilde{\Gm}^{A*}_{11}|<1/\Ga_A,\quad |\widetilde{\Gm}^{A*}_{1k}|<\Gb_A/\Ga_A,\quad
|\widetilde{\BGm}^{A*}_{k\ell}|<\Gb_A+\Gb_A^2/\Ga_A,
\eeq{1.15}
irrespective of the choice of $\Gl$.
Consequently, for any fixed choice of $\BP\ne 0$ and $\BQ$,
\beq  \sum_{i=1}^3\sum_{j=1}p_{ij}\widetilde{\Gve}^{A*}_{ij}\approx -\Gl{\rm Tr}(\BP^2)
\eeq{1.16}
approaches $-\infty$ as $\Gl\to\infty$, while
\beq \sum_{i=1}^3\sum_{j=1}^3q_{ij}\widetilde{\Gm}^{A*}_{ij}
\eeq{1.17}
remains bounded. It follows that $f_A(\BP,\BQ)$ is minus infinity unless $\BP=0$. Similarly
$f_B(\BP,\BQ)$ is zero unless $\BQ=0$ and thus $f(\BP,\BQ)$ is minus infinity unless both $\BP$ and $\BQ$
are zero. We conclude that any given pair of real symmetric matrices $({\BGve}^*,{\BGm}^*)$ is 
approximately realizable as the (effective permittivity, effective permeability) of a metamaterial.

\section*{Acknowledgements}
The author is grateful to Guy Bouchitt{\'e} for stimulating conversations
and is thankful for support from the
National Science Foundation through grant DMS-070978.

\bibliographystyle{mod-phaip}
\bibliography{/u/ma/milton/tcbook,/u/ma/milton/newref}

\ifx \bblindex \undefined \def \bblindex #1{} \fi\ifx \bblindex \undefined \def
  \bblindex #1{} \fi
\begin{thebibliography}{10}

\bibitem{MaxwellGarnett:1904:CMG}
J.~C. {Maxwell Garnett},
\newblock ``Colours in metal glasses and in metallic films'',
\newblock Philosophical Transactions of the Royal Society of London {\bf 203},
  385--420 (1904).

\bibitem{Veselago:1967:ESS}
V.~G. Veselago,
\newblock ``The electrodynamics of substances with simultaneously negative
  values of $\epsilon$ and $\mu$'',
\newblock Uspekhi Fizicheskikh Nauk {\bf 92}, 517--526 (1967),
\newblock English translation in {\em Soviet Physics Uspekhi} 10:509--514
  (1968).

\bibitem{Nicorovici:1994:ODP}
N.~A. Nicorovici, R.~C. McPhedran, and G.~W. Milton,
\newblock ``Optical and dielectric properties of partially resonant
  composites'',
\newblock Physical Review B (Solid State) {\bf 49}, 8479--8482 (1994).

\bibitem{Pendry:2000:NRM}
J.~B. Pendry,
\newblock ``Negative refraction makes a perfect lens'',
\newblock Physical Review Letters {\bf 85}, 3966--3969 (2000).

\bibitem{Cummer:2003:SCS}
S.~A. Cummer,
\newblock ``Simulated causal subwavelength focusing by a negative refractive
  index slab'',
\newblock Applied Physics Letters {\bf 82}, 1503--1505 (2003).

\bibitem{Shvets:2003:PAM}
G.~Shvets,
\newblock ``Photonic approach to making a material with a negative index of
  refraction'',
\newblock Physical Review B {\bf 67}, 035109 (2003).

\bibitem{Merlin:2004:ASA}
R.~Merlin,
\newblock ``Analytical solution of the almost-perfect-lens problem'',
\newblock Applied Physics Letters {\bf 84}, 1290--1292 (2004).

\bibitem{Podolskiy:2005:NSS}
V.~A. Podolskiy and E.~E. Narimanov,
\newblock ``Near-sighted superlens'',
\newblock Optics Letters {\bf 30}, 75--77 (2005).

\bibitem{Milton:2005:PSQ}
G.~W. Milton, N.-A.~P. Nicorovici, R.~C. McPhedran, and V.~A. Podolskiy,
\newblock ``A proof of superlensing in the quasistatic regime, and limitations
  of superlenses in this regime due to anomalous localized resonance'',
\newblock Proc. R. Soc. A {\bf 461}, 3999--4034 (2005).

\bibitem{Milton:2006:CEA}
G.~W. Milton and N.-A.~P. Nicorovici,
\newblock ``On the cloaking effects associated with anomalous localized
  resonance'',
\newblock Proc. R. Soc. A {\bf 462}, 3027--3059 (2006).

\bibitem{Nicorovici:2007:OCT}
N.-A.~P. Nicorovici, G.~W. Milton, R.~C. McPhedran, and L.~C. Botten,
\newblock ``Quasistatic cloaking of two-dimensional polarizable discrete
  systems by anomalous resonance'',
\newblock Optics Express {\bf 15}, 6314--6323 (2007).

\bibitem{Milton:2006:OPL}
G.~W. Milton, N.-A.~P. Nicorovici, and R.~C. McPhedran,
\newblock ``Opaque perfect lenses'',
\newblock Physica B {\bf 394}, 171--175 (2007).

\bibitem{Bruno:2007:SCS}
O.~P. Bruno and S.~Lintner,
\newblock ``Superlens-cloaking of small dielectric bodies in the quasistatic
  regime'',
\newblock Journal of Applied Physics {\bf 102}, 124502 (2007).

\bibitem{Bouchitte:2010:CSO}
G.~Bouchitt{\'e} and B.~Schweizer,
\newblock ``Cloaking of small objects by anomalous localized resonance'',
\newblock (2010),
\newblock Preprint,
  http://eldorado.tu-dortmund.de:8080/bitstream/2003/26457/1/mathematicalPrepr%
int14-09.pdf.

\bibitem{Lai:2009:CMI}
Y.~Lai, H.~Chen, Z.-Q. Zhang, and C.~T. Chan,
\newblock ``Complementary media invisibility cloak that cloaks objects at a
  distance outside the cloaking shell'',
\newblock Physical Review Letters {\bf 102}, 093901 (2009).

\bibitem{Kuhta:2007:FFI}
N.~A. Kuhta, V.~A. Podolskiy, and A.~L. Efros,
\newblock ``Far-field imaging by a planar lens: {Diffraction} versus
  superresolution'',
\newblock Physical Review B {\bf 76}, 205102 (2007).

\bibitem{Shelby:2001:EVN}
R.~A. Shelby, D.~R. Smith, and S.~Schultz,
\newblock ``Experimental verification of a negative index of refraction'',
\newblock Science {\bf 292}, 77--79 (2001).

\bibitem{Post:1962:FSE}
E.~J. Post,
\newblock {\em Formal structure of electromagnetics: {General} covariance and
  electromagnetics} (North-Holland Publishing Co., Amsterdam, 1962),
\newblock Dover Paperback, 1997.

\bibitem{Dolin:1961:PCT}
L.~S. Dolin,
\newblock ``To the possibility of comparison of three-dimensional
  electromagnetic systems with nonuniform anisotropic filling'',
\newblock Izv. Vyssh. Uchebn. Zaved. Radiofizika {\bf 4}, 964--967 (1961).

\bibitem{Pendry:2006:CEM}
J.~B. Pendry, D.~Schurig, and D.~R. Smith,
\newblock ``Controlling electromagnetic fields'',
\newblock Science {\bf 312}, 1780--1782 (2006).

\bibitem{Greenleaf:2003:ACC}
A.~Greenleaf, M.~Lassas, and G.~Uhlmann,
\newblock ``Anisotropic conductivities that cannot be detected by {EIT}'',
\newblock Physiological Measurement {\bf 24}, 413--419 (2003).

\bibitem{Greenleaf:2007:FWI}
A.~Greenleaf, Y.~Kurylev, M.~Lassas, and G.~Uhlmann,
\newblock ``Full-wave invisibility of active devices at all frequencies'',
\newblock Communications in Mathematical Physics {\bf 275}, 749--789 (2007).

\bibitem{Weder:2008:RAH}
R.~Weder,
\newblock ``A rigorous analysis of high-order electromagnetic invisibility
  cloaks'',
\newblock Journal of Physics A {\bf 41}, 065207 (2008).

\bibitem{Kohn:2009:CCV}
R.~V. Kohn, D.~Onofrei, M.~S. Vogelius, and M.~I. Weinstein,
\newblock ``Cloaking via change of variables for the helmholtz equation'',
\newblock Communications on Pure and Applied Mathematics (New York)  (2010),
\newblock to appear.

\bibitem{Schurig:2006:MEC}
D.~Schurig et~al.,
\newblock ``Metamaterial electromagnetic cloak at microwave frequencies'',
\newblock Science {\bf 314}, 977--980 (2006).

\bibitem{Cai:2007:OCM}
W.~Cai, U.~K. Chettiar, A.~V. Kildishev, and V.~M. Shalaev,
\newblock ``Optical cloaking with metamaterials'',
\newblock Nature Photonics {\bf 1}, 224--227 (2007).

\bibitem{Cai:2007:NMC}
W.~Cai, U.~K. Chettiar, A.~V. Kildishev, V.~M. Shalaev, and G.~W. Milton,
\newblock ``Non-magnetic cloak with minimized scattering'',
\newblock Applied Physics Letters {\bf 91}, 111105 (2007).

\bibitem{Li:2008:HUC}
J.~Li and J.~B. Pendry,
\newblock ``Hiding under the carpet: a new strategy for cloaking'',
\newblock Physical Review Letters {\bf 101}, 203901 (2008).

\bibitem{Liu:2009:BGP}
R.~Liu et~al.,
\newblock ``Broadband ground-plane cloak'',
\newblock Science {\bf 323}, 366--369 (2009).

\bibitem{Valentine:2009:OCD}
J.~Valentine, J.~Li, T.~Zentgraf, G.~Bartal, and X.~Zhang,
\newblock ``An optical cloak made of dielectrics'',
\newblock Nature Materials  (2009),
\newblock published online doi:10.1038/nmat2461.

\bibitem{Alu:2008:PMC}
A.~Al\'u and N.~Engheta,
\newblock ``Plasmonic and metamaterial cloaking: physical mechanisms and
  potentials'',
\newblock Journal of Optics A {\bf 10}, 093002 (2008).

\bibitem{Greenleaf:2009:CDE}
A.~Greenleaf, Y.~Kurylev, M.~Lassas, and G.~Uhlmann,
\newblock ``Cloaking devices, electromagnetic wormholes, and transformation
  optics'',
\newblock SIAM Review {\bf 51}, 3--33 (2009).

\bibitem{Chen:2007:TMR}
H.~Chen and C.~T. Chan,
\newblock ``Transformation media that rotate electromagnetic fields'',
\newblock Applied Physics Letters {\bf 90}, 241105 (2007).

\bibitem{Rahm:2008:DEC}
M.~Rahm et~al.,
\newblock ``Design of electromagnetic cloaks and concentrators using
  form-invariant coordinate transformations of maxwell's equations'',
\newblock Photonics and Nanostructures -- Fundamentals and Applications {\bf
  6}, 87--95 (2008).

\bibitem{Greenleaf:2008:EWV}
A.~Greenleaf, Y.~Kurylev, M.~Lassas, and G.~Uhlmann,
\newblock ``Electromagnetic wormholes and virtual magnetic monopoles from
  metamaterials'',
\newblock Physical Review Letters {\bf 99}, 183901 (2008).

\bibitem{Kildishev:2007:IMH}
A.~V. Kildishev and E.~E. Narimanov,
\newblock ``Impedance-matched hyperlens'',
\newblock Optics Letters {\bf 32}, 3432--3434 (2007).

\bibitem{Milton:2008:SFG}
G.~W. Milton, N.-A.~P. Nicorovici, R.~C. McPhedran, K.~Cherednichenko, and
  Z.~Jacob,
\newblock ``Solutions in folded geometries, and associated cloaking due to
  anomalous resonance'',
\newblock New Journal of Physics {\bf 10}, 115021 (2008).

\bibitem{Yan:2008:CSC}
M.~Yan, W.~Yan, and M.~Qiu,
\newblock ``Cylindrical superlens by a coordinate transformation'',
\newblock Physical Review B {\bf 78}, 125113 (2008).

\bibitem{Tsang:2008:MPL}
M.~Tsang and D.~Psaltis,
\newblock ``Magnifying perfect lens and superlens design by coordinate
  transformation'',
\newblock Physical Review B {\bf 77}, 035122 (2008).

\bibitem{Leonhardt:2006:OCM}
U.~Leonhardt,
\newblock ``Optical conformal mapping'',
\newblock Science {\bf 312}, 1777--1780 (2006).

\bibitem{Leonhardt:2009:BIN}
U.~Leonhardt and T.~Tyc,
\newblock ``Broadband invisibility by non-{Euclidean} cloaking'',
\newblock Science {\bf 323}, 110--112 (2009).

\bibitem{Tyc:2008:TSO}
T.~Tyc and U.~Leonhardt,
\newblock ``Transmutation of singularities in optical instruments'',
\newblock New Journal of Physics {\bf 10}, 115038 (2008).

\bibitem{Engheta:2005:CEO}
N.~Engheta, A.~Salandrino, and A.~{Al\'u},
\newblock ``Circuit elements at optical frequencies: {Nanoinductors},
  nanocapacitors, and nanoresistors'',
\newblock Physical Review Letters {\bf 95}, 095504 (2005).

\bibitem{Milton:2009:EC}
G.~W. Milton and P.~Seppecher,
\newblock ``Electromagnetic circuits'',
\newblock Networks and Heterogeneous Media  (2010),
\newblock Submitted, see also arXiv:0805.1079v2 [physics.class-ph] (2008).

\bibitem{Milton:2010:HEC}
G.~W. Milton and P.~Seppecher,
\newblock ``Hybrid electromagnetic circuits'',
\newblock Physica B  (2010),
\newblock Available online,doi:10.1016/j.physb.2010.01.007.

\bibitem{Silveirinha:2006:TEE}
M.~Silveirinha and N.~Engheta,
\newblock ``Tunneling of electromagnetic energy through subwavelength channels
  and bends using $\varepsilon$-near-zero materials'',
\newblock Physical Review Letters {\bf 97}, 157403 (2006).

\bibitem{Cai:2010:OM}
W.~Cai and V.~Shalaev,
\newblock {\em Optical Metamaterials: Fundamentals and Applications} (Springer,
  Dordrecht, 2010).

\bibitem{Efros:2004:DPC}
A.~L. Efros and A.~L. Pokrovsky,
\newblock ``Dielectric photonic crystal as medium with negative electric
  permittivity and magnetic permeability'',
\newblock Solid State Communications {\bf 129}, 643--647 (2004).

\bibitem{Bouchitte:2010:HFM}
G.~Bouchitt{\'e} and C.~Bourel,
\newblock ``Homogenization of finite metallic fibers and {3D}-effective
  permittivity tensor'',
\newblock Communications in Computational Physics  (2010),
\newblock To appear.

\bibitem{Pendry:1996:ELF}
J.~B. Pendry, A.~J. Holden, W.~J. Stewart, and I.~Youngs,
\newblock ``Extremely low frequency plasmons in metallic mesostructures'',
\newblock Physical Review Letters {\bf 76}, 4773--4776 (1996).

\bibitem{Bouchitte:1997:HSP}
G.~Bouchitt{\'e} and D.~Felbacq,
\newblock ``Homogenization of a set of parallel fibers'',
\newblock Waves in random media {\bf 7}, 1--12 (1997).

\bibitem{Bouchitte:2006:HWP}
G.~Bouchitt{\'e} and D.~Felbacq,
\newblock ``Homogenization of a wire photonic crystal: the case of small volume
  fraction'',
\newblock SIAM Journal on Applied Mathematics {\bf 66}, 2061--2084 (2006).

\bibitem{Schelkunoff:1952:ATP}
S.~A. Schelkunoff and H.~T. Friis,
\newblock {\em Antennas: the theory and practice}, pages 584--585,
\newblock John Wiley and Sons, New York~/ London~/ Sydney, Australia, 1952.

\bibitem{Pendry:1999:MCE}
J.~Pendry, A.~J. Holden, D.~J. Robbins, and W.~J. Stewart,
\newblock ``Magnetism from conductors and enhanced nonlinear phenomena'',
\newblock IEEE Transactions on Microwave Theory and Techniques {\bf 47},
  2075--2084 (1999).

\bibitem{Bouchitte:2010:HME}
G.~Bouchitt{\'e} and B.~Schweizer,
\newblock ``Homogenization of {Maxwell's} equations with split rings'',
\newblock SIAM Journal on Multiscale Modeling and Simulation  (2010),
\newblock To appear.

\bibitem{Zhikov:2000:EAT}
V.~V. Zhikov,
\newblock ``On an extension and an application of the two-scale convergence
  method'',
\newblock Matematicheskii Sbornik {\bf 191}, 31--72 (2000),
\newblock English translation in Sbornik: Mathematics, 191 (7), 973–1014
  (2000).

\bibitem{Zhikov:2004:GSE}
V.~V. Zhikov,
\newblock ``Gaps in the spectrum of some elliptic operators in divergent form
  with periodic coefficients'',
\newblock Algebra i Analiz {\bf 16}, 34--58 (2004),
\newblock English translation in St. Petersburg Mathematical Journal, 16 (5),
  773--790 (2005).

\bibitem{Bouchitte:2004:HNR}
G.~Bouchitt{\'e} and D.~Felbacq,
\newblock ``Homogenization near resonances and artificial magnetism from
  dielectrics'',
\newblock Comptes Rendus des S{\'e}ances de l'Acad{\'e}mie des Sciences.
  S{\'e}rie I. Math{\'e}matique {\bf 339}, 377--382 (2004).

\bibitem{Felbacq:2005:TMM}
D.~Felbacq and G.~Bouchitt{\'e},
\newblock ``Theory of mesoscopic magnetism in photonic crystals'',
\newblock Physical Review Letters {\bf 94}, 183902 (2005).

\bibitem{Kohn:2008:MHM}
R.~V. Kohn and S.~P. Shipman,
\newblock ``Magnetism and homogenization of microresonators'',
\newblock SIAM Journal on Multiscale Modeling and Simulation {\bf 7}, 62--92
  (2008).

\bibitem{Smyshlyaev:2009:PLE}
V.~P. Smyshlyaev,
\newblock ``Propagation and localization of elastic waves in highly anisotropic
  periodic composites via two-scale homogenization'',
\newblock Mechanics of Materials {\bf 41}, 434--447 (2009).

\bibitem{Maxwell:1954:TEM}
J.~C. Maxwell,
\newblock {\em A Treatise on Electricity and Magnetism}, volume~1, pages
  371--372,
\newblock Clarendon Press, Oxford, United Kingdom, 1873,
\newblock Article 322.

\bibitem{Milton:2002:TOC}
G.~W. Milton,
\newblock {\em The Theory of Composites} (volume~6 of {\em Cambridge Monographs
  on Applied and Computational Mathematics}Cambridge University Press,
  Cambridge, United Kingdom, 2002).

\bibitem{Tartar:1979:ECH}
L.~Tartar,
\newblock Estimation de coefficients homog{\'e}n{\'e}is{\'e}s. ({French})
  [{Estimation} of homogenization coefficients],
\newblock in {\em Computing Methods in Applied Sciences and Engineering: Third
  International Symposium, Versailles, France, December 5--9, 1977,}, edited by
  R.~Glowinski and J.-L. Lions, volume 704 of {\em Lecture Notes in
  Mathematics}, pages 364--373, Berlin~/ Heidelberg~/ London~/ etc., 1979,
  Springer-Verlag,
\newblock English translation in {\em Topics in the Mathematical Modelling of
  Composite Materials}, pp. 9--20, ed. by A. Cherkaev and R. Kohn. ISBN
  0-8176-3662-5.

\bibitem{Backus:1962:LWE}
G.~E. Backus,
\newblock ``Long-wave elastic anisotropy produced by horizontal layering'',
\newblock Journal of Geophysical Research {\bf 67}, 4427--4440 (1962).

\bibitem{Wiener:1912:TMF}
O.~Wiener,
\newblock ``{Die Theorie des Mischk{\"o}rpers f{\"u}r das Feld des
  station{\"a}ren Str{\"o}mung. Erste Abhandlung die Mittelswerts{\"a}tze
  f{\"u}r Kraft, Polarisation und Energie}\bblindex{Wiener bounds}. ({German})
  [{The} theory of composites for the field of steady flow. {First} treatment
  of mean value estimates for force, polarization and energy]'',
\newblock Abhandlungen der mathematisch-physischen Klasse der K{\"o}niglich
  S{\"a}chisischen Gesellschaft der Wissenschaften {\bf 32}, 509--604 (1912).

\end{thebibliography}

\end{document}